\documentclass[showpacs,amsmath,amssymb,aps,prd,preprint,groupedaddress,superscriptaddress]{revtex4-1} 
\usepackage[latin1]{inputenc}
\usepackage{graphicx}

\begin{document}


\title{Wormholes in Wyman's solution}


\author{J. B. Formiga}
\email[]{jansen.formiga@uespi.br}
\affiliation{Centro de Ciências da Natureza, Universidade Estadual do Piauí, C. Postal 381, 64002-150 Teresina, Piauí, Brazil}

\author{T. S. Almeida}
\email{talmeida@fisica.ufpb.br}

\affiliation{Universidade Federal da Paraíba, Departamento de Física, C. Postal 5008, 58051-970 João Pessoa, Pb, Brazil}
\date{\today}


\date{\today}

\begin{abstract}
The most general solution of the Einstein field equations coupled with a massless scalar field is known as Wyman's solution. This solution is also present in the Brans-Dicke theory and, due to its importance, it  has been studied in detail by many authors. However, this solutions has not been studied from the perspective of a possible wormhole. In this paper, we perform a detailed analysis of this issue. It turns out that there is a wormhole. Although we prove that the so-called throat cannot be traversed by human beings, it can be traversed by particles and bodies that can last long enough.
\end{abstract}


\maketitle

\newcommand{\aproxmenor}{ \put(2,0){$<$} \put(2,-5){$\sim$}\quad \ }
\newcommand{\aproxmaior}{ \put(2,0){$>$} \put(2,-5){$\sim$}\quad \ }

\section{Introduction}
Wormholes are one of the most intriguing objects that are allowed by the Einstein field equations. Theoretically, if they exist, they could perhaps be used as a shortcut to the furthest distances of our universe, connect our universe to another  one or even be used to time travel \cite{PhysRevLett.61.1446,PhysRevD.41.1116,Lorentzianwormholes}. There is no empirical evidence to support them yet and they have always been associated with exotic matter (matter that violates the energy conditions). Nevertheless, a lot of attention has been paid to their geometrical properties  and it is believed that quantum mechanics could provide such an exotic matter, since in the Casimir effect the null, weak, strong, and dominant energy conditions are all violated \footnote{See, for instance, Sec. 12.3.2 of Ref. \cite{Lorentzianwormholes}.}. Another empirical fact that supports this kind of exotic matter is the accelerated expansion of the universe, which can be explained by a matter that violates at least one of the energy conditions \cite{Nojiri20031,Straumann:2002he}. Some methods have been developed to distinguish the gravitational lensing due to  a wormhole from the ones caused by other objects \cite{PhysRevD.51.3117,*Virbhadra:1998dy,*doi:10.1142/S021773230100398X,*PhysRevD.65.103004,*PhysRevD.74.024020,*doi:10.1142/S0217732308025498,*PhysRevD.77.124014}. In short, we can say that the possibility of having wormholes in our universe is a very important aspect of general relativity.

Wyman's solution, also known as Fisher-Janis-Newman-Winicour solution, corresponds to the most general spherically symmetric solution to the Einstein--massless-scalar-field equations \cite{PhysRevD.24.839,PhysRevD.86.084031}. It contains a particular case that can be seen as describing a spherical body  and  which is in agreement with the solar-system experiments \cite{PhysRevD.83.087502}. One also finds Wyman's solution in the context of Brans-Dicke theory as a special case of the Campanelli-Lousto solutions \cite{PhysRevD.86.084031,doi:10.1142/S0218271893000325,PhysRevD.51.2011}, in an alternative version of this theory \cite{PhysRevD.89.064047}, and even in a model with torsion and nonmetricity \cite{10.1007/s10773-014-2003-2}. Due to its importance, it  has been studied in detail by many authors \cite{PhysRevD.24.839,PhysRevD.83.087502,PhysRevD.51.2011,PhysRevD.31.1280,PhysRevLett.20.878,Roberts:1993re,PhysRevD.40.2564,PhysRevD.81.024035,Fisher:1948yn,:/content/aip/journal/jmp/46/6/10.1063/1.1920308,OliveiraNetoSousa2008}. Some of them have even called the attention to a possible wormhole solution present in a particular case of the Wyman solution \cite{:/content/aip/journal/jmp/46/6/10.1063/1.1920308,OliveiraNetoSousa2008,OliveiraNetoSousa2008}. However, despite the great interest in this solution, as far as we know, no detailed analysis of its possible wormholes has been made so far. In this paper, we try to fill this gap by proving that there exist wormholes in Wyman solution and also by studying the properties of its throat. Our analysis is based on the properties of traversable wormholes listed in Ref. \cite{citeulike:5196630} and also on the definitions present in Ref. \cite{PhysRevD.56.4745}.

We begin in Sec. \ref{29012014a} with a list of properties that a wormhole should possess in order to be traversable by humans, while in Sec. \ref{1032014a} we present Wyman's solution and some of its features. Section \ref{1032014b} is devoted to the analysis of a wormhole that does not satisfies all Morris-Thorne conditions \cite{citeulike:5196630}, but does satisfy Hochberg and Visser general definition of wormhole \cite{PhysRevD.56.4745}. In this section, we also prove that its throat separates two regions where the curvature tensor goes to zero as we walk away from the throat, at least for certain values of one of the parameters presented in Wyman's solution. In addition, the detailed analysis reveals that this throat cannot be traversed by humans, although it could be by something else that could last long enough.  The results of this paper are summarized in Sec. \ref{1032014d}.

\section{Traversable Wormholes}\label{29012014a}

To describe a spherically symmetric wormhole, it is convenient to write the metric in the form
\begin{equation}
ds^2=e^{2\Phi(R)}dt^2-dR^2/\left[1-b(R)/R \right]-R^2d\Omega^2, \label{10082013a}
\end{equation}
where $b$ is known as the shape function and $\Phi$ as the redshift function \cite{citeulike:5196630}. The orthonormal basis of reference frame of static observers are given by
\begin{equation}
e_{\hat{t}}=e^{-\Phi}\partial_t, \quad e_{\hat{R}}=(1-b/R)^{1/2}\partial_{R}, \quad e_{\hat{\theta}}=R^{-1}\partial_{\theta},\quad e_{\hat{\varphi}}=(r\sin\theta)^{-1}\partial_{\varphi}.\label{10082013b}
\end{equation}

The functions $\Phi$ and $b$ must satisfy some conditions in order for the spacetime (\ref{10082013a})  to have a wormhole that can be traversed by humans. A list with such conditions was given by Morris-Thorne in Ref. \cite{citeulike:5196630}. However, a more general definition of wormhole can be found in Ref. \cite{PhysRevD.56.4745}. The latter definition is much wider and include the former as a particular case, hence, we will stick to it. Nonetheless, we write down Morris-Thorne list below so that we can cite each of these conditions properly. Of course, we have made some changes to adapt this list to the purpose of this paper.

\begin{center} List of properties of a human-traversable wormhole \footnote{ Adapted with permission from M. S. Morris and K. S. Thorne, American Journal of Physics 56, 395 (1988). \copyright 1988, American Association of Physics Teachers. }  \end{center}
\begin{enumerate}
\item Constraints on $b$ and $\Phi$:
	\begin{enumerate} \item General constraints: \label{03032014b}
		\begin{enumerate}    \item Spatial geometry is that of a wormhole.\label{03032014a} \item Throat is at minimum of $R$, denoted by  $R_m$ [in this case, we have $R_m=b_m\equiv b(R_m)$]; \label{9082013a} 
											                                    \item We must also have $1-b/R \geq 0$  everywhere; \label{9082013b} 
											                                    \item As $l \to \pm \infty$ we have $b/R \to 0$, where $l=\pm $ (the proper radial distance from wormhole throat as measured by the static observers). \label{9082013cc} 																													
											\item No horizons or singularities, i.e., $\Phi$ is finite everywhere; \label{9082013c} 
											\item $t$ measures proper time in asymptotically flat regions $\Leftrightarrow$ $\Phi \to 0$ as $l \to \pm \infty$. \label{9082013d}
											\end{enumerate}
										\item Description and constraints of a trip through wormhole  ($v$ is the radial velocity of traveler as measured by static observers, and $\gamma\equiv [1-(v/c)^2]^{-1/2}$; $c$ is the speed of light): 
		\begin{enumerate} 
											\item trip begins at $l=-l_1$ with $v=0$ and ends at $l=l_2$ with the same speed; \label{9082013dd}
											\item gravity is weak at $-l_1$ and $l_2$, that is, at these points
												\begin{enumerate} \item $b/R \ll 1$,\label{9082013e} \item $|\Phi|\ll 1$,\label{9082013f} 
																					\item $|\Phi'c^2| \aproxmenor g$, where $'\equiv d/dR$ and $g=\textrm{(Earth gravity)}$.\label{9082013g} 
																					\end{enumerate} 
											\item Trip takes less than one year from the point of view of both traveler and static observers at $-l_1$ and $l_2$. As a result, we must have
												\begin{equation}  \Delta\tau = \int_{-l_1}^{l_2}(v\gamma)^{-1}dl \aproxmenor 1\ yr, \label{9082013h} \end{equation}  \begin{equation} \Delta t = \int_{-l_1}^{l_2}(ve^{\Phi})^{-1}dl \aproxmenor 1\ yr. \label{9082013i} \end{equation}  
										  \item Traveler feels ``less'' than $g$ acceleration,
											\begin{equation} |e^{-\Phi} d(\gamma e^{\Phi})/dl| \aproxmenor g/c^2. \label{9082013j} \end{equation} 											
											\item Tidal-gravity accelerations between different parts of traveler's body is less than or approximately equal to $ g$:                                				\begin{equation}  \left|\left(1-b/R \right)\left[ -\Phi^{\prime\prime}+\frac{1}{2}\frac{(b'-b/R)}{R-b}\Phi'-(\Phi')^2 \right] \right|\aproxmenor 1/(10^{10}\ \textrm{cm})^2; \label{9082013l} \end{equation}
																				\begin{equation} \left|\frac{\gamma^2}{2R^2} \left[ \frac{v^2}{c^2} (b'-b/R)+2(R-b)\Phi' \right] \right|\aproxmenor 1/(10^{10}\ \textrm{cm})^2. \label{9082013m} \end{equation}
																																							 
											\item Traveler must not couple strongly to material that generates wormhole curvature.\label{9082013n}
											\end{enumerate}   
										\end{enumerate} \item Properties of the material that generates wormhole curvature: \label{11092013a} 
																			\begin{enumerate} \item Stress-energy tensor as measured by static observers: 
																				\begin{enumerate} \item \begin{eqnarray*} T_{\hat{t}\hat{t}}=\rho c^2=\textrm{(density of mass-energy)},\quad T_{\hat{R}\hat{R}}=-\tau= -\textrm{(radial tension)},
\\ T_{\hat{\theta}\hat{\theta}}=T_{\hat{\varphi}\hat{\varphi}}=p=\textrm{(lateral pressure)}. \end{eqnarray*} \label{9082013o} 
                                    \item Einstein field equations: \begin{eqnarray} \rho=\frac{b'}{8\pi Gc^{-2} R^2},\quad \tau=\frac{b/R-2(R-b)\Phi'}{8\pi Gc^{-4}R^2},
\nonumber \\ p=\frac{R}{2}\left[(\rho c^2-\tau)\Phi'-\tau' \right]-\tau. \label{9082013p}\end{eqnarray} In the throat ($R=R_m$), we have $\tau \approx 5\times 10^{41}\ dyn\ cm^{-2}(10\ m/b_m)^2$. 
																		\end{enumerate} \item (Field equations)+(absence of horizon at throat) $\Rightarrow$ $\tau>\rho c^2$ in throat $\Rightarrow$ traveler moving through throat at very high speed sees negative mass-energy density $\Rightarrow$ violation of weak, strong, and dominant energy conditions in throat. \label{9082013q} 
																										\item One might wish to require $\rho \geq 0$ everywhere (static observers see nonnegative mass-energy density), which implies $b'\geq 0$ everywhere.  \label{9082013r}
									  \end{enumerate} 
\end{enumerate} 

In Ref.~\cite{PhysRevD.56.4745}, Hochberg and Visser define a traversable wormhole throat as the two-dimensional hypersurface of minimal area taken in one of the constant-time spatial slices. When the minimal value of the area can be found by extremizing the area, one can show that the trace of the extrinsic curvature $K_{ab}$ of this two-surface vanishes. Besides, its derivative with respect to the normal coordinate $n$ (in Gaussian normal coordinates) is negative when one uses the definition    
\begin{equation}
K_{ab}=-\frac{1}{2}\frac{\partial g_{ab}}{\partial n}. \label{26062014b}
\end{equation}
 
This definition of wormhole encompasses the Morris-Thorne one, which is limited to two asymptotically flat regions that are spherically symmetric.  

\section{Wyman's solution}\label{1032014a}
Wyman's solution corresponds to the spherically-symmetric solution of the following Einstein's field equations
\begin{eqnarray}
G_{\mu \nu}=-\mu \left( V_{,\mu}V_{,\nu}-\frac{1}{2}g_{\mu \nu}V_{,\lambda}V^{,\lambda} \right),
\\
\Box V=0,
\end{eqnarray}
where $V$ is a scalar field and $\mu$ is the coupling constant. This solution can be written in the form
\begin{eqnarray}
ds^2=W^Sdt^2-W^{-S}dr^2-r^2W^{1-S}d\Omega^2, \label{10082013c}
\\
V=-\frac{1}{2\eta} \ln W, \label{10082013d}
\\
W=1-r_0/r,\qquad r_0=2\eta,\qquad \eta=\sqrt{M^2+\mu/2},\qquad S=M/\eta, \label{10082013e}
\end{eqnarray}
where $M$ is a constant and $d\Omega^2$ is the metric on a unit $2$-sphere. For $M$ and $\mu$ positive, we have a spacetime that has a naked singularity  and  can be thought of as representing the exterior region of a spherical body of mass $M$. The case $S=1$ corresponds to the Schwarzschild spacetime. In this paper, however, we shall deal only with the case $M>0$ and $-2M^2<\mu<0$ (the same as $S>1$).

To write the metric (\ref{10082013c}) in the form given by Eq. (\ref{10082013a}), we just need to compare Eqs. (\ref{10082013c})-(\ref{10082013e}) with Eq. (\ref{10082013a}). This comparison shows that
\begin{eqnarray}
R=rW^{(1-S)/2}, \label{10082013f}
\\
\Phi(R)=\frac{S}{2} \ln W(r(R)), \label{10082013g}
\\
b/R=1-\frac{1}{W(r(R))}\left[1-(1+S)\frac{r_0}{2r(R)} \right]^2, \label{10082013h}
\end{eqnarray}

\subsubsection{the minimum of $R$}
From Eq. (\ref{10082013f}), one finds that the minimum value of $R$ occurs at
\begin{equation}
r_m=\frac{S+1}{2}r_0, \label{10082013hh}
\end{equation}
which yields
\begin{equation}
R_m=r_m \left(\frac{S-1}{S+1} \right)^{(1-S)/2}. \label{10082013i}
\end{equation}
It is clear that the relation between the radial coordinates $R$ and $r$ is one-to-one only for certain values of $r$, which depends on the possible values of $S$. For $S\geq 1$, this relation is one-to-one for $r\in [r_m,\infty)$ and the values of $R$ are those in the interval $[R_m,\infty)$. If we take $S<1$, the values of $R$ will be $(0,\infty)$.  Nonetheless, unlike $R$, the domain of $r$ is always $ (r_0,\infty)$.
As we shall see later, there is a throat at $r_m$ that ``separates'' the regions $(r_0,r_m)$ and $(r_m,\infty)$.

\section{The wormhole in Wyman's solution} \label{1032014b}

\subsection{Coordinate $w$}\label{20022014a}
In dealing with wormholes, it is sometimes interesting to work with a coordinate that does not posses coordinate singularities. Although $r$
is not singular in the interval $(r_0,\infty)$, let us define a coordinate $w$ analogous to $l$ through the integral
\begin{equation}
w=\int_{r_m}^{r} W^{-S/2}dr, \label{18122013a}
\end{equation}
where it is clear that $w=0$ corresponds to $r_m$ (the throat). We call $A$ the region with negative values of $w$ and $B$ the other region.

Since the analytic solution of (\ref{18122013a}) for an arbitrary $S$ may not exist, consider the following expansion for the integrand.
\begin{eqnarray}
(1-r_0/r)^{-S/2}= 1+\frac{Sr_0}{2}\frac{1}{r}+  \sum_{n=2}^{\infty} \frac{r_0^n}{2^n n!}r^{-n}\prod_{j=1}^{n}(S+2j-2). \label{15092013d}
\end{eqnarray}
By substituting the expansion (\ref{15092013d}) into the integral (\ref{18122013a}), we obtain the expression
\begin{eqnarray}
w= r-r_m+\frac{Sr_0}{2}\ln(r/r_m) +\sum_{n=2}^{\infty} \frac{r_0^n}{2^n n!(n-1)}\left(  r_m^{1-n}-r^{1-n} \right)   \prod_{j=1}^{n}(S+2j-2). \label{18122013b}
\end{eqnarray}
From the ratio test, one can easily prove that the series above converges for $r>r_0$ and $S>1$, but Raabe's test shows that it diverges at $r_0$  for $S>2$ (it converges for $1<S<2$); from the integral (\ref{18122013a}), we see that $w$ is infinite at $r_0$ for $S=2$. 

\subsubsection{Choosing a value for $w_1$} \label{23122013e}
Like $l_1$ and $l_2$, $w_1$ and $w_2$ will represent the places where the trip begins and ends, respectively. The ideal value for $w_1$ is the one that favors the conditions listed in Sec. \ref{29012014a}. For reasons that will become clear in other sections (see, for instance, Sec. \ref{23122013a}), we choose $r_1$ as
\begin{equation}
r_1=M(1+S)^2/(2S^2). \label{29012014b}
\end{equation}
One can easily check that this point is between $r_0$ and $r_m$ for $S>1$, which is the case we are interested in.

\begin{figure}[h]
\includegraphics[scale=.43]{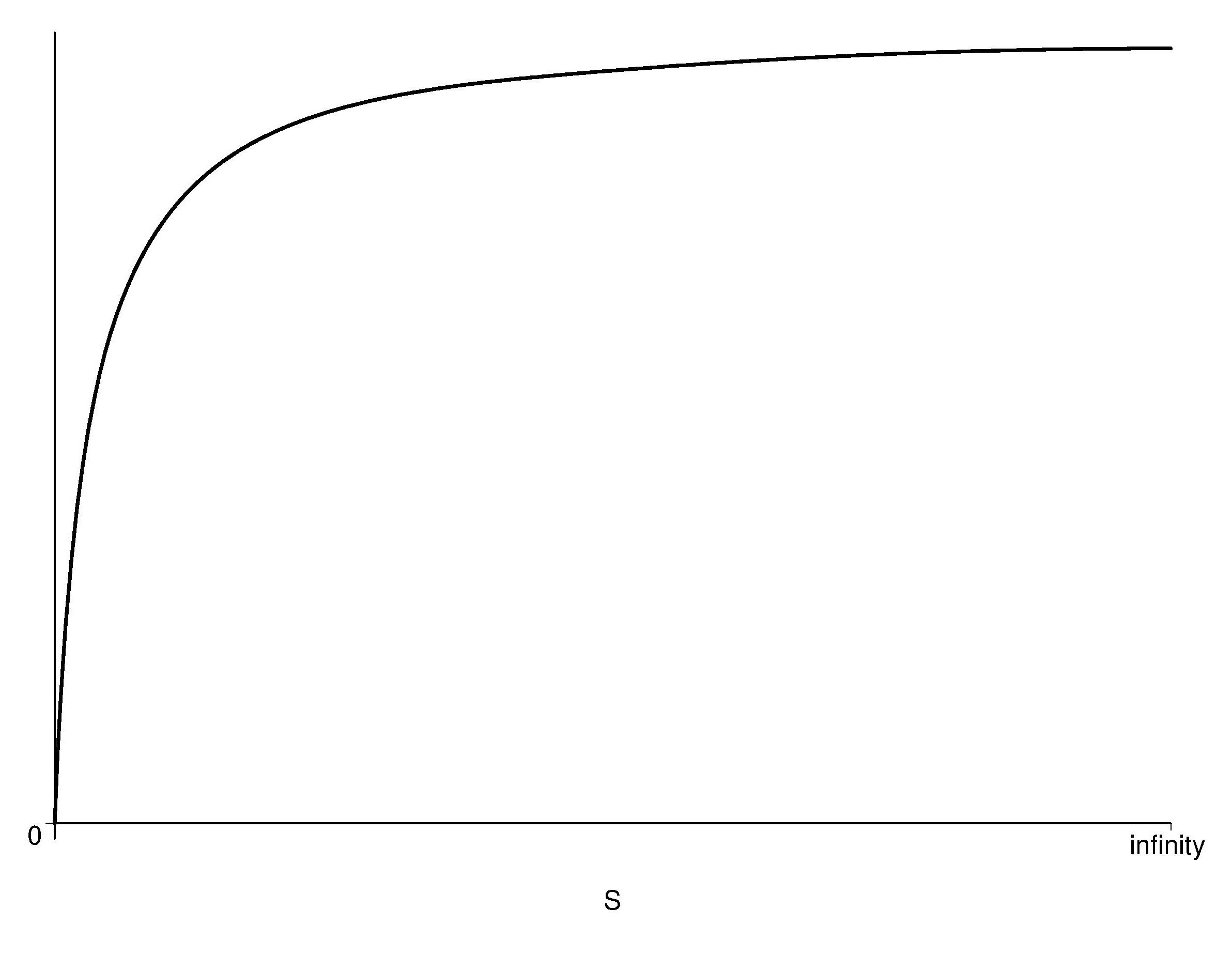}
\caption{\scriptsize This figure shows a plot of $w_1$ as a function of $S$ from $1$ to $\infty$, where we have set $M=1$ and used one hundred terms of the series in Eq. (\ref{22122013a}). One can verify that the qualitative behavior of the curve will not change if we increase the number of terms. }
\label{22012014a}
\end{figure}

Defining $w_1$ as $-w(r_1)$ and using Eq. (\ref{18122013b}), we get
\begin{eqnarray}
w_1=M\biggl\{ (S^2-1)/(2S^2)+\ln\left(\frac{2S}{1+S} \right) +\sum_{n=2}^{\infty} \frac{[2S/(1+S)]^{n-1}-1}{(n-1)n!S(1+S)^{n-1}} \prod_{j=1}^{n}(S+2j-2) \biggr\}. \label{22122013a}
\end{eqnarray}
From Fig. \ref{22012014a}, which is a plot of $w_1$ as a function of $S$ from $1$ to $\infty$, we see that $w_1$ is a monotonically increasing function of $S$ with its minimum at $S=1$ (note that $w_1=0$ for $S=1$). From Eq. (\ref{22122013a}), one can evaluate the maximum value of $w_1$ (take the limit $S\to \infty$) and find that  $0<w_1\aproxmenor 2M$.

\subsubsection{Choosing a value for $w_2$}\label{20022014b}
From the metric (\ref{10082013c}), it is clear that $r_2$ has to be big enough to decrease significantly  the values of terms like $M/r$. However, it cannot be too big because of the conditions (\ref{9082013h}) and (\ref{9082013i}). To accommodate these requirements, we use $r_2=10^AM$ with $A\aproxmaior 6$ (for more details about this constraint, see Sec. \ref{23122013a}). By substituting $r_2$ in Eq. (\ref{18122013b}), one finds that  $w_2\approx r_2$, which together with the maximum value of $r_1$ (see  Sec. \ref{23122013e}) leads to $w_1+w_2\approx w_2$. This result will be used later.

\subsection{The behavior of the curvature far from the throat} \label{04022014b}
From Eq. (8) in Ref. \cite{citeulike:5196630}, we see that the only nonvanishing components of the Riemann tensor in the basis (\ref{10082013b}) have the form
\begin{eqnarray}
Form_1= \left(1-b/R \right)\left[ -\Phi^{\prime\prime}+\frac{1}{2}\frac{(b'-b/R)}{R-b}\Phi'-(\Phi')^2 \right], \label{23112013a} \\
\nonumber \\
Form_2=-\frac{(1-b/R)}{R}\Phi', \label{23112013b}\\
\nonumber \\
Form_3=\frac{b'R-b}{2R^3} \label{23112013c}\\
\nonumber \\
Form_4=\frac{b}{R^3}. \label{23112013d}
\end{eqnarray}
From Eqs. (\ref{10082013f})-(\ref{10082013h}), one can evaluate Eq. (\ref{23112013a}) at $r_0$ to obtain
\begin{equation}
Form_1(r_0)=\frac{S}{r_0^2}[1-(1+S)/2]\lim_{r\to r_0}(1-r_0/r)^{S-2}=\Biggl\{ \begin{array}{lr} 0 & S>2,\\ \textrm{finite}\neq 0 & S=2,\\ \infty & 1<S<2. \end{array} \label{23112013e}
\end{equation}
The remaining expressions take the form
\begin{equation}
Form_2=-\frac{W_m^2}{W}\frac{\Phi'}{R}=-\frac{r_0SW_m}{2r^3}W^{S-2}, \label{23112013f}
\end{equation}
\begin{equation}
Form_3=-\frac{Sr_0}{2r^3}\left[1-\frac{(1+S)}{2S}\frac{r_0}{r} \right]W^{S-2}, \label{23112013h}
\end{equation}
\begin{equation}
Form_4=\frac{W-W_m^2}{r^2}W^{S-2},\label{23112013j}
\end{equation}
where we are using $W_m=1-r_m/r$. It is straightforward to check that the ``forms'' (\ref{23112013f})-(\ref{23112013j}) yield the same qualitative result as that of Eq. (\ref{23112013e}). Therefore,  the region $A$ becomes flat far from the throat only for $S>2$. Here, we call the attention to the fact that the region $A$ may be bounded, i.e., the time to go from $r_m$ to $r_0$ from the viewpoint of the traveler may be finite. If the point $r_0$ is not a physical singularity, then one will have to maximally extend Wyman's manifold to see what happens below $r_0$. 

\subsection{Condition \ref{9082013a}}
The two-surface characterized by a fixed moment of time and $\theta=\pi/2$ cannot be completely embedded into a three-dimensional Euclidean space. We can see that in the following way. From Eq. (27) in Ref. \cite{citeulike:5196630}, we have
\begin{equation}
\frac{dz}{dR}=\pm \left(R/b-1 \right)^{-1/2}, \label{30012014c}
\end{equation}
where $z$ is the ``$z-$coordinate'' of the cylindrical coordinate system. For $r<r_1$, the term $R/b-1$ is negative. Thus,   the interval $(r_0,r_1]$ cannot be used in Eq. (\ref{30012014c}). Nonetheless, we can embed the portion $(r_1,\infty)$ just to see how the two-surface looks like. In this case, Eq. (\ref{30012014c}) can be written in the form
\begin{equation}
\frac{dz}{dR}=\frac{\sqrt{W-W_m^2}}{W_m}\quad \Rightarrow \quad  \frac{dz}{dr}=(W-W_m^2)^{1/2}W^{-(1+S)/2}, \label{24062014n}
\end{equation}
where we have used the chain rule and $\pm |W_m|=W_m$ (the negative values of $W_m$ represent the region $z<0$).

Due to the cylindrical symmetry, we can parametrize the two-surface $t=constant$ and $\theta=\pi/2$ as
\begin{equation}
\chi(r,\phi)=(R(r),\phi,z(R(r))). \label{26062014a}
\end{equation}
The solution of Eq. (\ref{24062014n}) will give us the explicit form of $\chi$.

\begin{figure}[h]
\includegraphics[scale=.43]{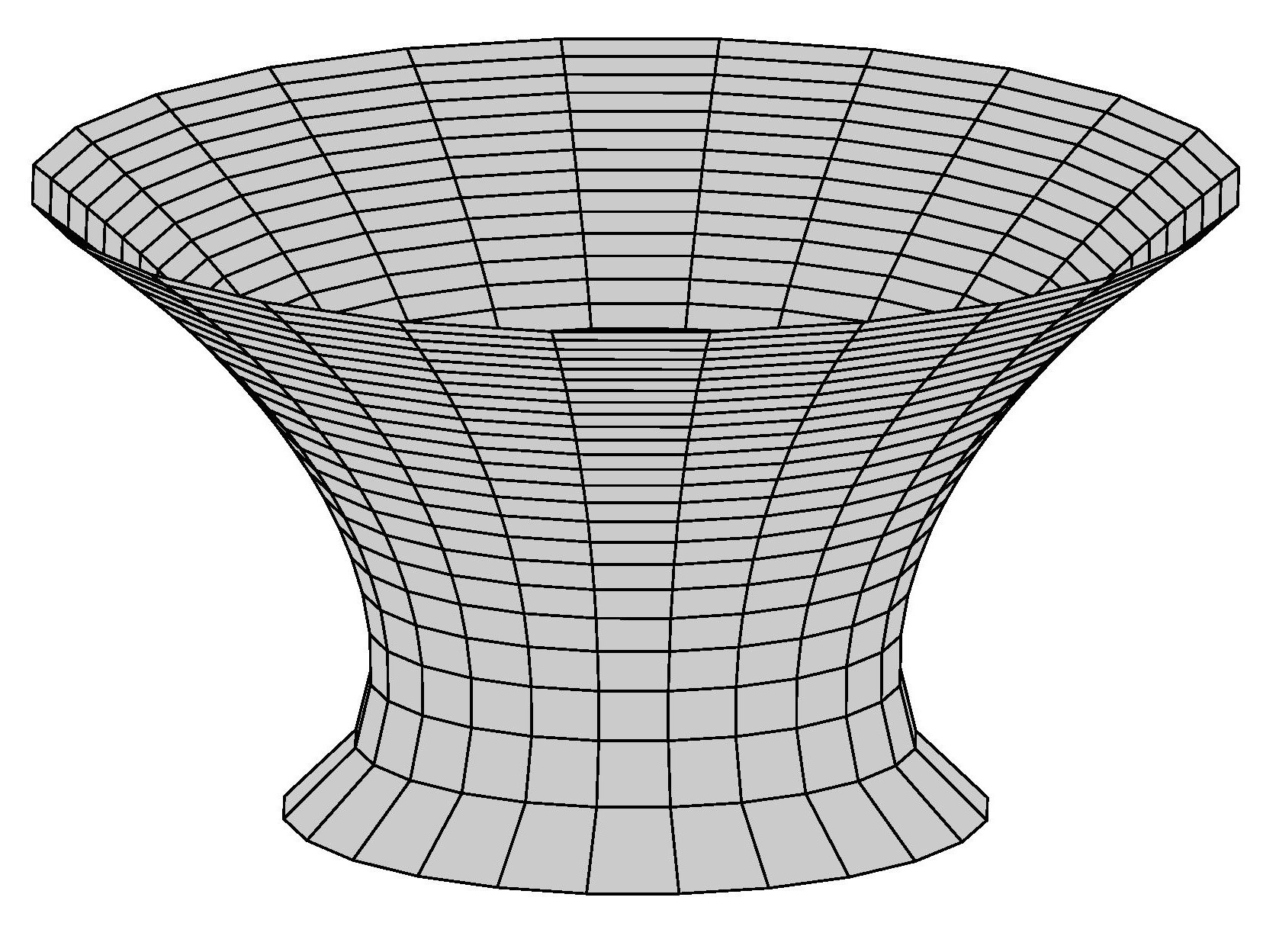}
\caption{\scriptsize In this figure we exhibit the plot of the two-surface (\ref{26062014a}) with $r_0=1$ and $r$ varying from $4/3$ to $8$. As one can see, this surface has the shape of a typical wormhole. }
\label{24062014r}
\end{figure}

For $S=3$, we manage to obtain the following exact solution:
\begin{equation}
z=r_0\arctan \left(\frac{\sqrt{3r_0r-4r^2_0}}{r_0}\right)+\frac{2r-3r_0}{r-r_0}\sqrt{3r_0r-4r^2_0}-\left(\arctan\sqrt{2}+\sqrt{2} \right)r_0, \label{24062014p}
\end{equation}
where the constant of integration has been chosen in such a way that the throat is at $z=0$. By using Eqs. (\ref{10082013f}) and (\ref{24062014p}) in Eq. (\ref{26062014a}), one obtains the plot in Fig.~\ref{24062014r}. 

With the help of a computer, we have verified that the case $S=3/2$ yields a two-surface similar to the one in Fig.~\ref{24062014r}. This case is different from the previous one because $r_0$ is a physical singularity for $1<S<2$ (no need for maximal extension). 

\subsection{Conditions \ref{9082013b}, \ref{9082013cc}, \ref{9082013c} and \ref{9082013d}} \label{19022014a}
From Eq. (\ref{10082013h}) and the fact that $r\in (r_0,\infty)$, we have $1-b/R\geq 0$ everywhere.

The condition $w \to \infty \Rightarrow b/R \to 0$ is clearly satisfied, since $b/R \to 0$ as $r\to \infty$. However, when $w\to -\infty$, which is equivalent to $r\to r_0$, we have $b/R \to -\infty$ for $S>1$ [see Eq. (\ref{10082013h})]. Hence, the condition \ref{9082013cc} is not satisfied.

It is evident from Eq. (\ref{10082013g}) that $\Phi$ is finite for $r>r_0$.

From Eq. (\ref{10082013g}), we see that the condition \ref{9082013d} is not satisfied because $w\to -\infty$  (the same as $r\to r_0$) implies $\Phi\to -\infty$.

It is clear that the object that we are studying is not a Morris-Thorne wormhole. Nevertheless, we are going to prove now that it is indeed a wormhole. The candidate to be the throat is the two-surface $t=constant$, $R=constant$. Applying these constraints to Eq. (\ref{10082013a}), one gets
\begin{equation}
ds^2_2=-R^2(d\theta^2+\sin^2\theta d\phi^2) \label{27062014a}
\end{equation}
and 
\begin{equation}
\partial_{n_{\pm}}=\pm \sqrt{1-b/R}\ \partial_R, \label{27062014b}
\end{equation}
where $\partial_{n_{\pm}}$ is the unit normal vector with plus sign  for $r>r_m$ and the minus one for $r_0<r<r_m$. Finally, using Eqs. (\ref{27062014a}) and (\ref{27062014b}) into Eq. (\ref{26062014b}), we arrive at
\begin{equation}
tr(K)=\mp 2\frac{\sqrt{1-b/R}}{R}, \label{24062014h}
\end{equation}
where $tr(K)$ stands for $g^{ab}K_{ab}$. From Eqs. (\ref{10082013h}) and (\ref{10082013hh}), we see that the trace of the extrinsic curvature vanishes at $r_m$. In turn, by applying $\partial_{n_{\pm}}$ to Eq. (\ref{24062014h}), one obtains
\begin{equation}
\frac{\partial tr(K)}{\partial n_{\pm}}=\frac{2}{R^2} \left[\frac{R}{2}\frac{d}{dR}bR^{-1}+1-b/R \right]. \label{24062014i}
\end{equation}
Evaluating this expression at $r_m$ for the Wyman metric [see Eqs.~(\ref{10082013f}) and (\ref{10082013h})], we find that
\begin{equation}
\frac{\partial tr(K)}{\partial n_{\pm}}\Biggl|_{r_m}=-\frac{2}{r_mR_m}(1-r_0/r_m)^{(S-1)/2}<0. \label{24062014j}
\end{equation}
We have now proved that the ``strong flare-out condition'' is satisfied and, therefore, we have a wormhole in Wyman spacetime for $S>1$.

\subsection{Conditions \ref{9082013dd}, \ref{9082013e}, \ref{9082013f},  \ref{9082013g}, (\ref{9082013h}), (\ref{9082013i}), (\ref{9082013j}), (\ref{9082013l}), e (\ref{9082013m})} \label{13122013a}
Now we analyze the conditions that are ``necessary'' to ensure that the wormhole is traversable by humans.

\subsubsection{Conditions \ref{9082013dd}, \ref{9082013e}, \ref{9082013f},  \ref{9082013g}} \label{23122013a}
The condition \ref{9082013dd} is just a matter of convenience, hence it is not a problem. With respect to the conditions \ref{9082013e}, \ref{9082013f} and \ref{9082013g}, we can assume that $r_0/r_2=2M/(Sr_2) \ll 1$ with $w_2=w_2(r_2)$ being the place where the trip ends. From Eqs. (\ref{10082013g}) and (\ref{10082013h}), one gets $|\Phi|\approx M/r_2$  and $b/R\approx 2M/r_2$,  where all these values have been evaluated at $r_2$. Deriving Eq. (\ref{10082013g}) with respect to $R$, one finds that 
\begin{equation}
\Phi'=\frac{r_0S}{2r^2} \frac{W^{(S-1)/2}}{W_m}, \label{15082013c} 
\end{equation}
which yields the approximation $\Phi'\approx M/r^2_2$ (Remember that we have defined $W_m=1-r_m/r$). If we use the constraint \ref{9082013g} in this approximation, we will obtain
\begin{equation}
 r_2 \aproxmaior c\sqrt{\frac{M}{g}},  \label{15082013f}
\end{equation}
Writing $r_2$ in the form $r_2=10^AM$ and substituting it into Eq. (\ref{15082013f}),  one finds that $A\aproxmaior \log(c/\sqrt{Mg} )$. On the other hand, if we impose the conditions \ref{9082013e} and \ref{9082013f}, we will have $10^{-A} \ll 1$. Based on the latter inequality, it is reasonable to take $A\aproxmaior 6$. It can be shown that $\log(c/\sqrt{Mg} )$ is larger than $6$  only for $M< 9171\ m$, which allows us to take $A\aproxmaior 6$ whenever $M\aproxmaior 9171\ m$. 

With respect to $b/R\ll 1$ evaluated at $w=-w_1$, we have to be very careful because in this case the function $b/R$ does not decrease as $w$ goes to $-\infty$ ($r\to r_0$). In fact, it diverges there. Nonetheless, this function vanishes at $r_1$. That is the reason why we are using $w_1=-w(r_1)$ with $r_1$ given by Eq. (\ref{29012014b}).

Now we show that the condition \ref{9082013f} is not satisfied. From Eqs. (\ref{29012014b}) and (\ref{10082013g}), we get 
\begin{equation}
\Phi(r_1)=S\ln\left(\frac{S-1}{S+1} \right). \label{27122013e}
\end{equation}
The minimum value of $|\Phi(r_1)|$ is $2$ and occurs when $S$ goes to infinity. So we cannot have $|\Phi|\ll 1$ at $r_1$. Nevertheless, this result does not seem to be a real problem because the traveler feels force, not potential. The conditions that are related to forces are given by \ref{9082013g}, (\ref{9082013j}), (\ref{9082013l}), and (\ref{9082013m}). As we will see later, there are values for which these conditions can be satisfied.  

Since $\Phi'\approx M/r^2_2=10^{-2A}/M$, the constraint  $|\Phi'(r_2)|\aproxmenor g/c^2$  is weaker than $|\Phi'(r_1)|\aproxmenor g/c^2$. This means that the possible values for $M$ have to be taken from the latter inequality. The substitution of $r_1$ into Eq. (\ref{15082013c}) leads to
\begin{equation}
|\Phi'(r_1)|=\frac{4S^4}{M(1+S)^4} \left( \frac{S-1}{S+1}  \right)^{S-2}, \label{15082013d}
\end{equation}
whose minimum occurs when $S$ goes to infinity. Taking this limit ( $M$ is fixed), we obtain
\begin{equation}
\lim_{S\to \infty}|\Phi'(r_1)|=\frac{4}{Me^2}. \label{15082013e}
\end{equation}
Using this in the inequality \ref{9082013g}, we find that $M\aproxmaior 5\times 10^{15}\ m$. This is clearly a very large value for $M$, at least if we think of $M$ as being the mass of some spherically symmetric distribution of matter. It is worth mentioning that the above limit is equivalent to taking $\mu \to -2M^2$, that is, we must have $\mu\aproxmenor -5\times 10^{31}\ m$.

\subsubsection{Conditions (\ref{9082013h}) e  (\ref{9082013i})} \label{6012014a}
Let us assume that $v$ is constant. In this case, the condition (\ref{9082013h}) can be rewritten as
\begin{equation}
\Delta \tau=\frac{\sqrt{1-v^2/c^2}}{v}10^AM \aproxmenor 1\ yr, \label{15122013d}
\end{equation}
where we have used $w_1+w_2\approx w_2\approx r_2=10^AM$. Notice that, here, we are using $w$ rather than $l$.

With respect to the condition (\ref{9082013i}), the constancy of $v$ leads to
\begin{equation}
\Delta t=\frac{1}{v}\int_{-w_1}^{w_2}W^{-S/2}dw \aproxmenor 1\ yr. \label{23122013b}
\end{equation}
Note that, due to the similarity between $l$ and $w$, we can simply exchange $l$ for $w$ in Eq. (\ref{9082013i}) to write this condition in terms of $w$. Using [see Eq. (\ref{18122013a})]
\begin{equation}
dw=W^{-S/2}dr, \label{31012014b}
\end{equation}
we find that
\begin{equation}
\Delta t=\frac{1}{v}\int_{r_1}^{r_2}W^{-S}dr. \label{23122013c}
\end{equation}
The integrand of this expression can be expanded in the form
\begin{equation}
(1-r_0/r)^{-S}=1+\frac{Sr_0}{r}+\sum_{n=2}^{\infty} \frac{r_0^n}{n!}r^{-n}\prod_{j=0}^{n-1}(S+j). \label{16092013h}
\end{equation}
Using this expansion in Eq. (\ref{23122013c}), we arrive at
\begin{equation}
\Delta t=\frac{1}{v}\left[ r_2-r_1+Sr_0\ln(r_2/r_1)+\sum_{n=2}^{\infty} \frac{r_0^n}{n!(1-n)}\left(r_2^{1-n}-r_1^{1-n}\right)\prod_{j=0}^{n-1}(S+j)  \right].	\label{23122013d}
\end{equation}
Since the largest value of $r_1$ is $2M$, while $r_2=10^AM$ with $A\aproxmaior 6$, we can approximate the above expression to 
\begin{equation}
\Delta t\approx \frac{r_2}{v}=\frac{10^AM}{v}.\label{23122013g}
\end{equation}
The smallest value for $M$ that is allowed by the condition \ref{9082013g} is   $M=5\times 10^{15}\ m$. Using this value in Eq. (\ref{23122013g}) with $A=6$ and taking $v$ as the speed of light, one gets $\Delta t \approx 5\times 10^5\ yr$. Since this is the best-case scenario, we can conclude from this result that it is not possible to satisfy the conditions \ref{9082013g} and (\ref{9082013i}) simultaneously. In Sec. \ref{31012014a}, we show that the conditions (\ref{9082013h}) and (\ref{9082013j}) cannot be satisfied simultaneously either.

\subsection{Condition (\ref{9082013j})}
The assumption that $v$ is constant allows us to rewrite the inequality (\ref{9082013j}) in the form
\begin{equation}
\gamma \left| \frac{d\Phi}{dr} \frac{dr}{dw} \right| \aproxmenor g/c^2, \label{26122013a}
\end{equation}
where, from now on, we denote the left-hand side of this inequality by $f$.

\begin{figure}[h]
\includegraphics[scale=.44]{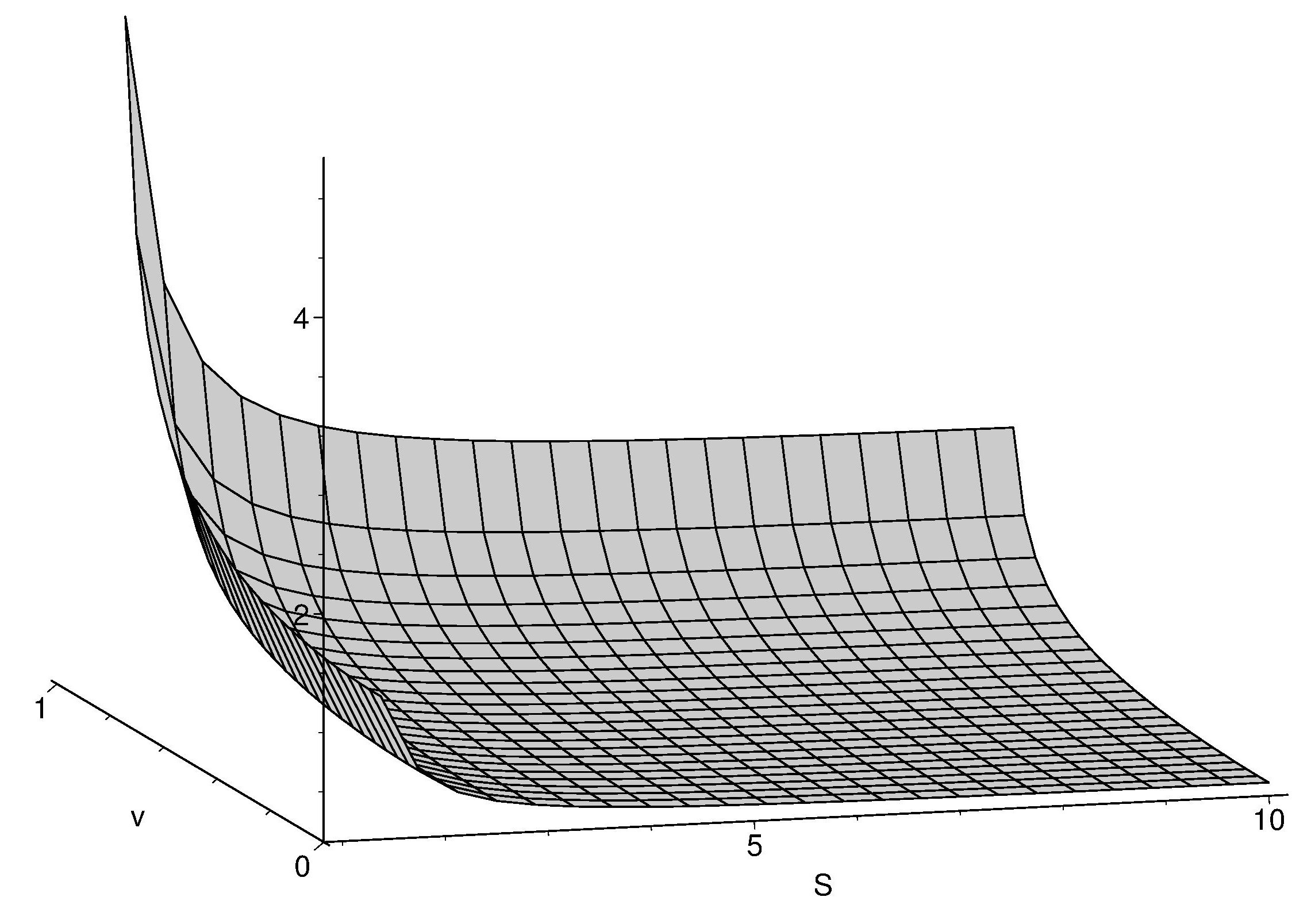}
\caption{\scriptsize This figure shows the behavior of $f(r_1)$ as a function of $S$ and $v$ in the intervals $S\in (1,10)$ and $v\in (0,1)$, where we have used $M=c=1$. It can be shown that the qualitative behavior of the above surface does not change for larger values of $S$.}
\label{09012014a}
\end{figure}

Substituting Eqs. (\ref{31012014b}) and (\ref{10082013g}) into Eq. (\ref{26122013a}) gives 
\begin{equation}
f= \gamma\frac{r_0S}{2r^2}W^{S/2-1} \aproxmenor g/c^2. \label{26122013b}
\end{equation}
To know the maximum value of $f$ during the trip, we need to calculate the local maximums and compare the respective values of $f$ evaluated at these points with the values $f(r_1)$ and $f(r_2)$. If the above inequality holds for the maximum value of $f$, then it holds for any other value. A simple calculation shows that there is only one maximum for $f$ and it is given by
\begin{equation}
r_c=\frac{(2+S)r_0}{4}. \label{26122013c}
\end{equation}
At first glance we could consider this point to be relevant because it is in the domain of $r$ for $S\geq 2$, remember that $r\in(r_0,\infty)$. However, the traveler does not reach this point, since $r_c$ is less than $r_1$. Thus, we need to compare only $f(r_1)$ with $f(r_2)$. The value of $f$ at $r_1$ is [see Eq. (\ref{29012014b})]
\begin{equation}
f(r_1)=\frac{4}{M\sqrt{1-v^2/c^2}}\frac{S^4}{(S+1)^4} \left(  \frac{S-1}{S+1} \right)^{S-2}, \label{08012014c}
\end{equation}
where we have used $\gamma=(1-v^2/c^2)^{-1/2}$. By using $r_2=10^AM$ with $A\aproxmaior 6$, one can easily verifies that $f(r_2)<f(r_1)$.  Therefore, the condition (\ref{9082013j}) becomes
\begin{equation}
\frac{4}{M\sqrt{1-v^2/c^2}}\frac{S^4}{(S+1)^4} \left(  \frac{S-1}{S+1} \right)^{S-2} \aproxmenor g/c^2. \label{08012014d}
\end{equation}
A plot of $f(r_1)$ as a function of $S$ and $v$ is shown in Fig. \ref{09012014a}. From this plot, one can see that $f(r_1)$ reaches its minimum as $S$ goes to infinity.
In this limit, we have
\begin{equation}
\frac{4}{Me^2\sqrt{1-v^2/c^2}} \aproxmenor g/c^2. \label{08012014e}
\end{equation}
Note that this constraint is stronger than that imposed by condition \ref{9082013g} [see Eq. (\ref{15082013e})]. In addition, it requires low speed, which is not good for the inequalities (\ref{9082013h}) and (\ref{9082013i}). In what follows, we prove that the conditions (\ref{9082013h}) and (\ref{9082013j}) cannot be satisfied simultaneously.

\subsection{ The conflict between the inequalities (\ref{9082013h}) and (\ref{9082013j})} \label{31012014a}
It is clear in Eq. (\ref{15122013d}) that the condition (\ref{9082013h}) ask for high speed. However, from Eq. (\ref{08012014d}), we see that the condition (\ref{9082013j}) do exactly the opposite. Thus, the best case occurs when we take the largest value of $v$ allowed by Eq. (\ref{08012014d}). Taking the equality in Eq. (\ref{08012014d}), we find that
\begin{eqnarray}
\sqrt{1-v^2/c^2}=\frac{4c^2}{Mg}\frac{S^4}{(1+S)^4}\left[ (S-1)/(S+1) \right]^{S-2}, \label{27122013a}
\\
\nonumber \\
v=c\sqrt{1-B^2}, \qquad B^2=\frac{16c^4}{M^2g^2}\frac{S^8}{(1+S)^8}\left[ (S-1)/(S+1) \right]^{2S-4}. \label{27122013b}
\end{eqnarray}

By using Eqs. (\ref{27122013a}) and  (\ref{27122013b}) into (\ref{15122013d}), one arrives at
\begin{equation}
\frac{4c10^A}{g\sqrt{1-B^2} }\frac{S^4}{(1+S)^4}\left[ (S-1)/(S+1) \right]^{S-2} \aproxmenor 1\ yr. \label{27122013c}
\end{equation}
In the most favorable case, i.e., $A=6$, $S\to \infty$, and $B\to 0$ ( $M\to \infty$), the left-hand side of this inequality becomes
\begin{equation}
 \frac{4c}{g e^2}\times 10^6\approx 5\times 10^5\ yr. \label{27122013d}
\end{equation}
This result is clearly in contradiction with Eq. (\ref{27122013c}). Thus, we conclude that the conditions (\ref{9082013h}) and (\ref{9082013j}) cannot be satisfied simultaneously.

\subsection{Condition (\ref{9082013l})}\label{22032014a}
After some calculations, we find that the left-hand side of the inequality (\ref{9082013l}) for the metric (\ref{10082013c}) can be written as
\begin{equation}
f_1=\frac{Sr_0}{r^3}W_mW^{S-2}. \label{07032014a}
\end{equation}
One can show that $f_1$ possesses two critical points, which are given by
\begin{equation}
 r_{\pm}=\left(\frac{1+S}{2} \pm \frac{\sqrt{3S^2-3}}{6} \right)r_0. \label{22082013g}
\end{equation}
A simple calculation shows that $r_+$ is in the interval $[r_1,r_2]$, but $r_-$ is not. Thus, the maximum value of $f_1$ can occur only at  $|f_1(r_1)|$, $|f_1(r_+)|$ or $|f_1(r_2)|$. 

\begin{figure}[h]
\includegraphics[scale=.41]{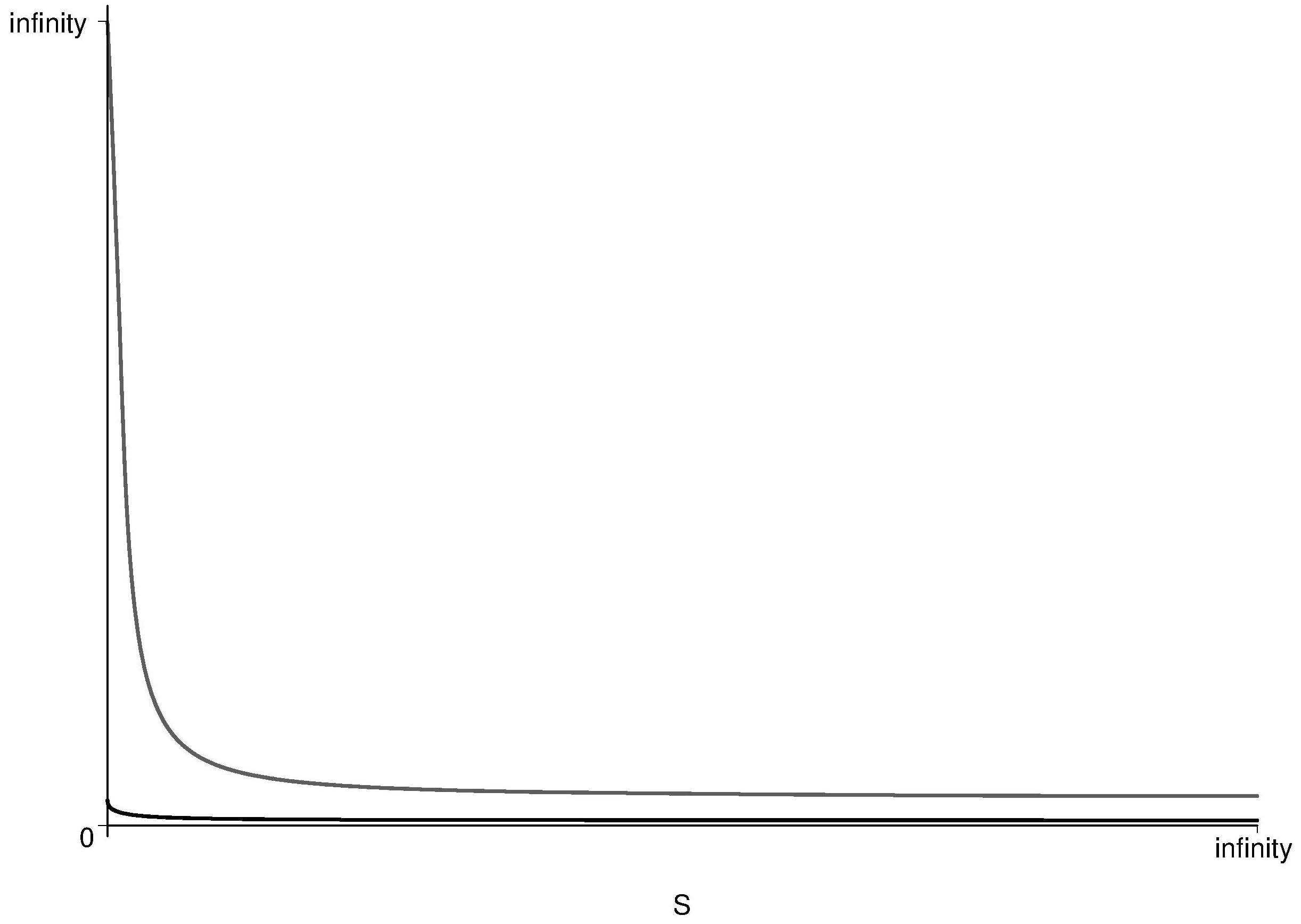}
\caption{\scriptsize In this figure the black curve represents $|f_1(r_+)|$ as a function of $S$, while the grey one is the plot of $|f_1(r_1)|$. The interval between $0$ and $1$ has been suppressed so that the qualitative behavior of these curves in the interval $(1,\infty)$ be better visualized. }
\label{3022014a}
\end{figure}

Calculating $|f(r_1)|$, one finds that
\begin{equation}
|f_1(r_1)|=\frac{16}{M^2}\frac{S^6 (S^2-1)}{(1+S)^8} \left(  \frac{S-1}{S+1} \right)^{2S-4}. \label{08012014f}
\end{equation}
It is easy to check that $f_1(r_2)\approx 2\times 10^{-3A}/M^2 < |f_1(r_1)|$. By comparing $|f_1(r_1)|$ with $|f_1(r_+)|$, we also find that $|f(r_1)|$ is always bigger (see Figs. \ref{3022014a}). Hence, $|f(r_1)|$ corresponds to the largest value of the left-hand side of the inequality (\ref{9082013l}) during the trip. 

From the inequality (\ref{9082013l}) we see that $|f_1(r_1)|\aproxmenor 10^{-16}/m^2$. As shown in Fig. \ref{3022014a}, the minimum value of $|f_1(r_1)|$ occurs when $S\to \infty$.  Therefore, this limit is our best choice for $S$. By taking this limit, we get
\begin{equation}
\frac{16}{M^2e^4} \aproxmenor 10^{-16}/m^2,  \label{08012014h}
\end{equation}
which yields $M\aproxmaior 5\times 10^7\ $m.

\subsection{Condition (\ref{9082013m})} \label{08032014b}
\begin{figure}[h]
\includegraphics[scale=.44]{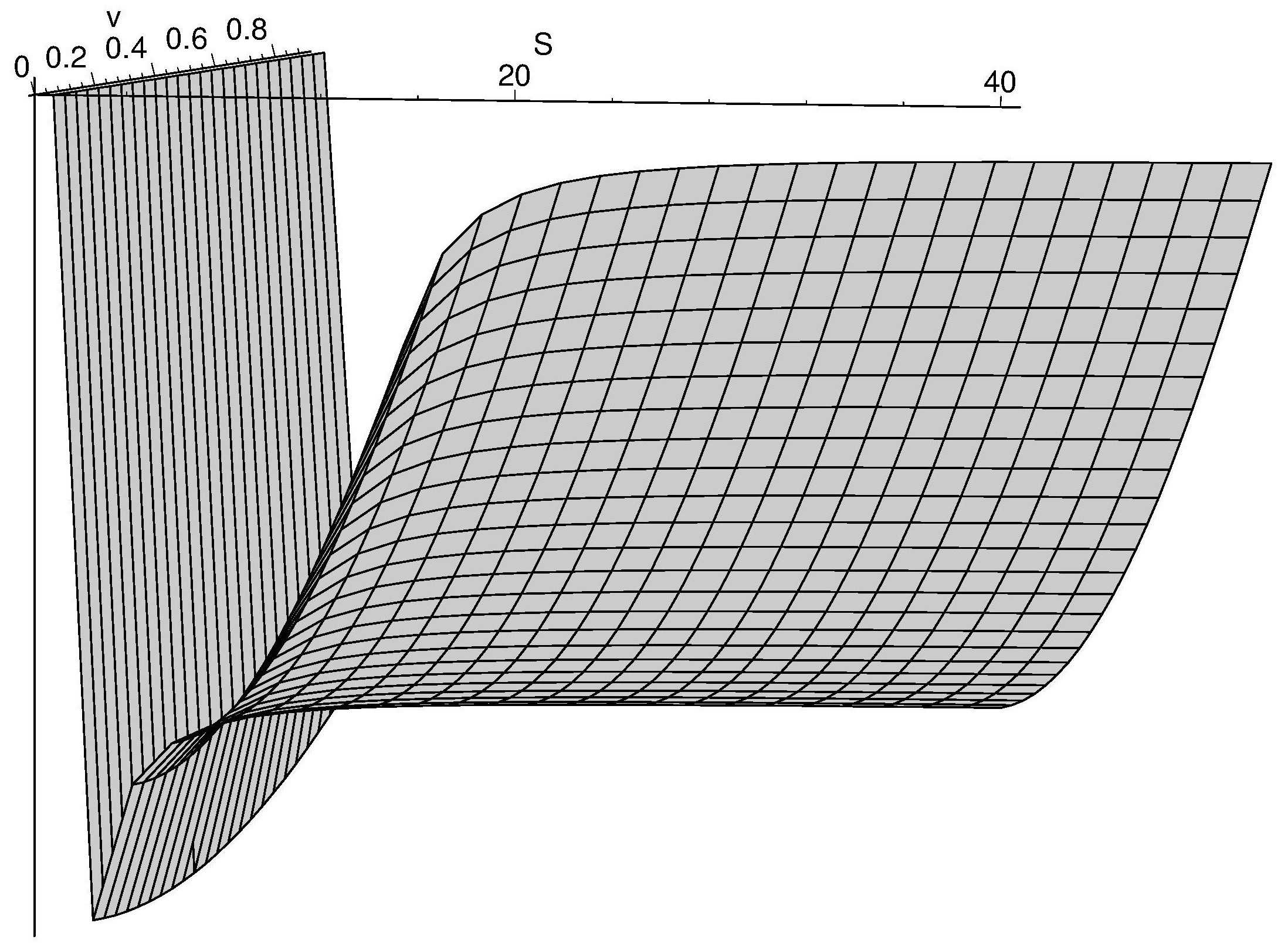}
\caption{\scriptsize In this figure we see the plot of $r_--r_1$ as a function of $S$ and $v$, where we have used $M=c=1$. The above surface has only negative values, which suggests that $r_-<r_1$. One can verify that this qualitative behavior will not change if we increase the range of values of $S$. In this plot, we have used the interval $[0,0.9]$ for $v$. The reason why we have not used the value $v=1$ is because the program used to make this plot is not able to properly evaluate $r_-$ at this point. Nevertheless, there is no divergence there. One can verify that the limit of $r_-$ as $v$ goes to $1$ and $S$ is kept fixed is $(S+2)M/(2S)$.}
\label{09012014c}
\end{figure}

Treating the left-hand side of the inequality (\ref{9082013m}) as a function of $r$, which we denote by $f_2$, we find that
\begin{equation}
f_2=\left|\frac{Sr_0}{2r^3}W^{S-2} \left[ 1-\frac{r_0}{2r}-\frac{Sr_0}{2r}\left(1-\frac{v^2}{S^2c^2} \right)\gamma^2 \right]\right|. \label{29082013c}
\end{equation}
For $v$ constant, there are two critical points, namely,
\begin{equation}
r_{\pm}=\left\{ \frac{(S+1)[ 3S-(S+2)v^2/c^2 ]  \pm \sqrt{(S^2-1)[3S^2+(S^2-4)v^4/c^4]} }{6S} \right\}r_0\gamma^2. \label{29082013h}
\end{equation}
As we can see from Fig. \ref{09012014c}, the point $r_-$ is outside the interval $[r_1,r_2]$.
Thus, we are left with  $|f_2(r_1)|$, $|f_2(r_+)|$, and $|f_2(r_2)|$. 

\begin{figure}[h]
\includegraphics[scale=.44]{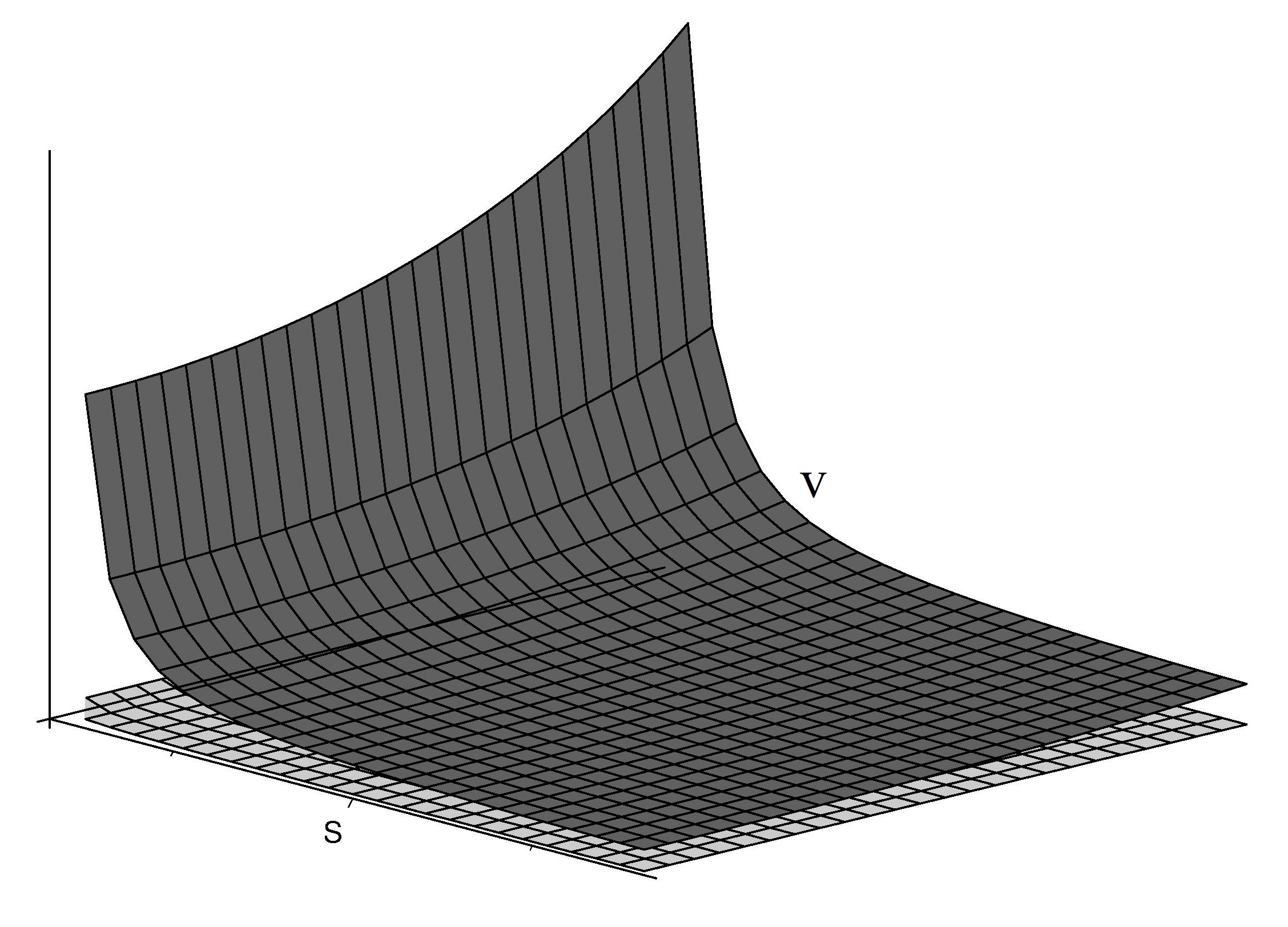}
\caption{\scriptsize  
The black surface corresponds to the plot of $|f_2(r_1)|$, which has been treated as a function of $S$ and $v$. The other surface corresponds to $|f_2(r_+)|$. In this figure, we have set $c=M=1$, $v\in [0,0.5]$, and $S\in [1,3]$. As this figure suggests, we have $|f_2(r_1)|>|f_2(r_+)|$. The qualitative behavior of the surfaces does not change for a wider range of values of $v$ and $S$.}
\label{09012014d}
\end{figure}

Evaluating  $f_2(r_1)$, we find that
\begin{equation}
f_2(r_1)=\frac{8}{M^2}\frac{S^6}{(1+S)^6} \left( \frac{S-1}{S+1} \right)^{2S-4} \left[ 1-\frac{2S}{(1+S)^2}-\frac{2S^2}{(1+S)^2}\frac{1-v^2/(Sc)^2}{1-v^2/c^2} \right].   \label{08012014i}
\end{equation}
By comparing this with $|f_2(r_+)|$ (see Fig. \ref{09012014d}), we see that $ |f_2(r_1)|>|f_2(r_+)|$.
In addition, we also have $|f_2(r_2)| <|f_2(r_1)|$ \footnote{One can see this last result by plotting  $|f_2(r_1)|$ and $|f_2(r_2)|$ with $A=6$. Since this plot is very similar to figure \ref{09012014d}, we will not show it here.}.  Therefore, the maximum value of the left-hand side of Eq. (\ref{9082013m}) during the trip occurs at $r_1$ and is given by (\ref{08012014i}).

A natural question we may ask ourselves is what the values of $S$ and $v$ that minimize $|f_2(r_1)|$ are. Figure \ref{09012014d} can be used to answer this question. As it suggests, the minimum of $|f_2(r_1)|$ happens when $S \to \infty$ and $v\to 0$. Substituting these values in Eq. (\ref{08012014i}), one finds that
\begin{equation}
\frac{8}{M^2e^4} \aproxmenor 10^{-16}/m^2,   \label{08012014j}
\end{equation}
which yields $M\aproxmaior 4\times 10^7\ m$. This is basically the same result yielded by the condition (\ref{9082013l}).

\subsection{Condition \ref{9082013n}}
The condition \ref{9082013n} is clearly problematic. If we image that $M$ is the mass of a body such as a star or a planet, then it is impossible to have a traversable wormhole. Nevertheless, we can still consider the possibility of having a different kind of matter that may meet the requirement \ref{9082013n}. The so-called dark matter, for example, would clearly satisfy this condition, since it does not couple strongly with ordinary matter.

\subsection{The matter distribution}
To have an idea of how the matter that generates the wormhole of the Wyman solution is distributed over space, let us see how the density of mass-energy behaves in the frame of the static observers. Using the metric (\ref{10082013c}) in $\rho$ as given by (\ref{9082013p}), one finds that
\begin{equation}
\rho=-\frac{S^2-1}{32\pi Gc^{-2}}\frac{r_0^2}{r^4}W^{S-2}, \label{5012014c}
\end{equation}
which is negative for $S>1$. This means that the static observers see negative mass-energy density. Therefore, Wyman's solution does not satisfies the requirement \ref{9082013r}.

Note that, for $S>2$, the mass-energy density goes to zero as $r$ goes to $r_0$. This is in agreement with the result of Sec. \ref{04022014b}.

\section{Final remarks}\label{1032014d}
We have shown that the Wyman solution contains wormholes for $S>1$ without using the cut-paste technique\footnote{In Ref.~\cite{Barcelo1999127}, the authors study a solution of the Einstein field equations that is conformally related to the spacetime (\ref{10082013c}) for $S \in [-1,1]$. They also find wormholes without the need of the cut-paste technique. It is likely that, by following a procedure similar to the one adopted by these authors, one should be able to obtain a solution that is conformally related to (\ref{10082013c}) for $S>1$ and find new wormholes, perhaps with new interesting properties.}. For $S>2$, the two regions of the wormholes become flat as we walk away from the throat. In this case and also for $S=2$, there is a possibility that $r_0$ is not a physical singularity and a maximal extension may be needed. On the other hand, for $1<S<2$ we have seen that $r_0$ is an essential singularity. In all cases, the wormholes cannot be traversed by humans because Eqs. \ref{9082013g} and (\ref{9082013i}) cannot hold simultaneously; the same goes for (\ref{9082013h}) and (\ref{9082013j}). This problem happens because the ``time conditions''  (\ref{9082013h}) and (\ref{9082013i}) require $M$ to be small, while practically all others require the opposite. If we are to abandon these two conditions  based on the assumption that time is not a problem, then we must have $M\aproxmaior 4c^2/(ge^2\sqrt{1-v^2/c^2})\geq 5\times 10^{15}\ m$, which is clearly a strong constraint on $M$. Nevertheless, these wormholes are traversable in the sense that their throats remain opened and can be traversed by anything that last long enough.

At this point one may ask why big values of $M$ have been good for the constraints, except the ``time conditions''. The answer to this question is simple. The best setting for most of the constraints happens when $S$ goes to infinity, which means $\mu \to -2M^2$. While $M>0$ favors attraction, the negative values of $\mu$ produces a repulsive force. The latter can be seen from Eq. (\ref{5012014c}) [keep in mind that $S>1 \Leftrightarrow \mu<0$; see, e.g., Eq. (\ref{10082013e})]. Therefore, in this limit, big values of $M$ implies much bigger values of $|\mu|$.
 
The question whether $r_0$ is a physical singularity for  $S\geq 2$ will be studied in the future. This study may lead to the conclusion that the throat connects two asymptotically flat regions for $S>2$.

\section*{Acknowledgments}

T. S. Almeida would like to thank CAPES for financial support. 


%

\end{document}